\documentclass[aps,pra,reprint,superscriptaddress]{revtex4-1}
\usepackage{graphicx}

\renewcommand{\figurename}{Fig.}
\newcommand{\figref}[1]{\figurename~\ref{#1}}

\let\vec\mathbf
\pdfoutput=1
\begin{document}
	
	\title{Vectorial vortex generation and phase singularities upon Brewster reflection}
	
	\date{\today}
	
	\author{Ren\'{e} Barczyk}
	\affiliation{Max Planck Institute for the Science of Light, Staudtstr. 2, D-91058 Erlangen, Germany}
	\affiliation{Institute of Optics, Information and Photonics, University Erlangen-Nuremberg, Staudtstr. 7/B2, D-91058 Erlangen, Germany}
	\author{Sergey Nechayev}
	\email{sergey.nechayev@mpl.mpg.de}
	\affiliation{Max Planck Institute for the Science of Light, Staudtstr. 2, D-91058 Erlangen, Germany}
	\affiliation{Institute of Optics, Information and Photonics, University Erlangen-Nuremberg, Staudtstr. 7/B2, D-91058 Erlangen, Germany}
	\author{Abdullah Butt}
	\affiliation{Max Planck Institute for the Science of Light, Staudtstr. 2, D-91058 Erlangen, Germany}
	\affiliation{Institute of Optics, Information and Photonics, University Erlangen-Nuremberg, Staudtstr. 7/B2, D-91058 Erlangen, Germany}
	\affiliation{School in Advanced Optical Technologies, University Erlangen-Nuremberg, Paul-Gordan-Str. 6, D-91052 Erlangen, Germany}
	\author{Gerd Leuchs}
	\affiliation{Max Planck Institute for the Science of Light, Staudtstr. 2, D-91058 Erlangen, Germany}
	\affiliation{Institute of Optics, Information and Photonics, University Erlangen-Nuremberg, Staudtstr. 7/B2, D-91058 Erlangen, Germany}
	\author{Peter Banzer}
	\affiliation{Max Planck Institute for the Science of Light, Staudtstr. 2, D-91058 Erlangen, Germany}
	\affiliation{Institute of Optics, Information and Photonics, University Erlangen-Nuremberg, Staudtstr. 7/B2, D-91058 Erlangen, Germany}
	
	\begin{abstract}
		\noindent We experimentally demonstrate the emergence of a purely azimuthally polarized vectorial vortex beam with a phase singularity upon Brewster reflection of focused circularly polarized light from a dielectric substrate. The effect originates from the polarizing properties of the Fresnel reflection coefficients described in Brewster's law. An astonishing consequence of this effect is that the reflected field's Cartesian components acquire local phase singularities at Brewster's angle. Our observations are crucial for polarization microscopy and open new avenues for the generation of exotic states of light based on spin-to-orbit coupling, without the need for sophisticated optical elements.
	\end{abstract}
	
	\maketitle
	
	\section{Introduction} 
	
	Apart from scalar wave properties like intensity and phase, light also has intrinsic spatial vectorial degrees of freedom, described by its polarization distribution \cite{born2013principles}. Recently, there is strongly increasing interest in the generation and characterization of complex polarization states \cite{rubinsztein2016roadmap,zhan2009cylindrical}. Their remarkable properties are of paramount importance for a broad range of applications, such as 3D focus shaping \cite{zhan2002focus}, laser-based material processing \cite{niziev1999influence}, tight focusing of light \cite{dorn2003sharper} and \AA ngstr\"om-scale position sensing \cite{neugebauer2016polarization,bag2018transverse}, to name a few.
	
	Akin to mechanical objects, light may also possess angular momentum (AM), composed of orbital (OAM) and spin (SAM) parts \cite{allen1992orbital,o2002intrinsic,yao2011orbital,bliokh2010angular,bliokh2015transverse}. While SAM is attributed to the vectorial (circular) polarization of light, OAM is associated with the distribution of the scalar phase of a beam, possessing a helical pattern in its cross section with a singularity of arbitrary integer topological charge $\ell$ on the beam axis. The study of optical OAM has received considerable attention in the literature \cite{rubinsztein2016roadmap,franke2008advances,bliokh2015spin}, displaying great potential in various disciplines, including optical manipulation \cite{he1995direct}, quantum information protocols \cite{stutz2007create} and microscopy \cite{hell2003toward}.
	
	The coupling between SAM and OAM of light via spin-orbit interaction (SOI) has been studied extensively in the past (see \cite{bliokh2015spin} and references therein), e.g. as a means of controlling the OAM or the direction of propagation of an optical beam by its polarization \cite{biener2002formation,bomzon2002space,bliokh2010angular,bliokh2011spin,marrucci2006optical,marrucci2011spin,brasselet2009optical,gorodetski2010plasmonic,karimi2014generating,garoli2016optical,shitrit2013spin,lin2013polarization}. A common route towards mediating SOI involves the use of sub-wavelength gratings \cite{bomzon2002space,biener2002formation} or anisotropic inhomogeneous media \cite{marrucci2006optical,brasselet2009optical,Ciattoni_uniaxial,Ciattoni_uniaxial_exp,tilted_uniax,brasselet_uniax}. Another recent approach is based on the polarizing properties of axicons \cite{fink1979polarization} and it utilizes metallic or dielectric conical reflectors, where spin-to-orbital angular momentum conversion originates from phase changes upon total internal reflection as well as from the spin-redirection geometric phase \cite{bisson2006radially,mansuripur2011spin,kobayashi2012helical,bouchard2014achromatic,aleksanyan2016spin,radwell2016achromatic}. All of these methods rely on the intrinsic or geometric properties of dedicated optical elements. However, SOI naturally occurs as a consequence of AM conservation upon focusing \cite{zhao2007spin,bliokh2010angular,bliokh2011spin,fernandez2012helicity,fernandez2013electromagnetic} - an inherent process in microscopy.\\
	\indent In this letter, we report on the emergence of vectorial vortex beams, bearing phase singularities, in a surprisingly elementary cylindrically symmetric experimental configuration. The effect is based on SOI of tightly focused circularly polarized (CP) fields \cite{bliokh2015spin,bliokh2010angular,bliokh2011spin,zhao2007spin}, reflected from an unstructured planar dielectric substrate \cite{novotny2001reflected,nasalski2001three,nasalski2006polarization,nasalski2005excitation,yavorsky2012polarization} at Brewster's angle \cite{brewster1815laws}. Even over 200 years after its first description, research articles dealing with Brewster's angle are still being published \cite{paniagua2016generalized}, reporting on remarkable observations such as an enhanced spin Hall effect of light \cite{luo2011enhanced,kong2012spin} and mode conversion upon Brewster reflection \cite{aiello2009brewster}. Here, we use Brewster's effect to obtain a vector beam with a polarization and phase vortex from incident tightly focused homogeneously CP light. We validate our findings by polarization sensitive measurements of the reflected fields' intensity and phase distribution. We analyze the  focusing objective's back focal plane (BFP) in the cylindrical transverse electric/transverse magnetic (TE/TM) as well as in the Cartesian X/Y polarization basis. For the cylindrical coordinate frame, a central phase vortex is present in the azimuthal (TE) component. In the case of Cartesian coordinates, we see two phase singularities appearing at Brewster's angle for projections of the reflected light in the BFP onto the X and Y axis, respectively. In consequence, we prove the inevitable presence of polarization and phase vortices for any high numerical aperture ($\mathrm{NA}$) focusing geometry covering Brewster's angle, rendering our observations important especially in the field of polarization microscopy \cite{empedocles1999three,sonnichsen2005gold,whittaker1994quantitative}.     
	
	\begin{figure*}[!t]
		\centering 
		\includegraphics[width=\textwidth]{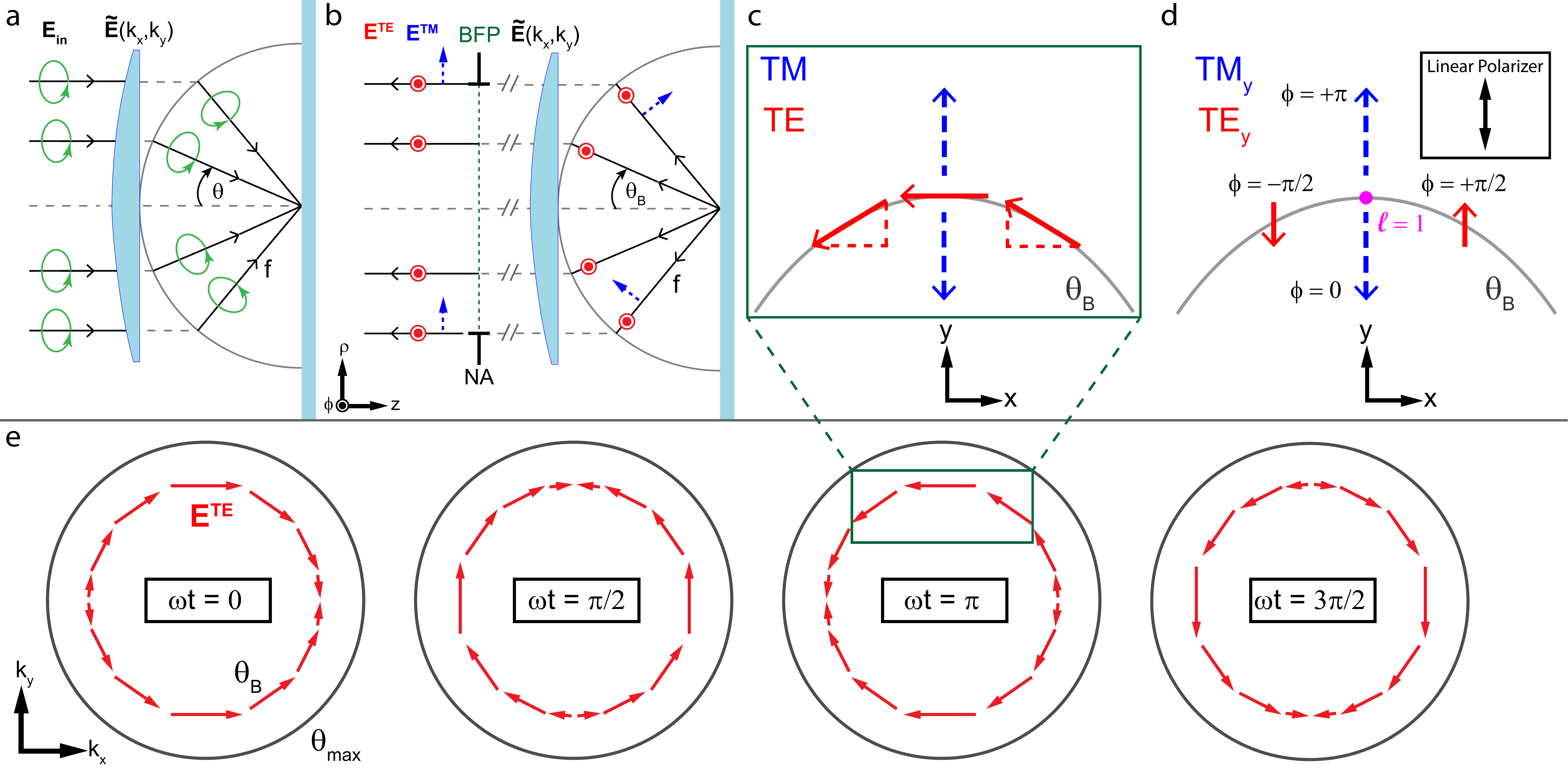} 
		\caption{(\textbf{a}) Focusing scheme of an aplanatic lens for incident circular polarization (CP). The paraxial input field $\vec{E}_{\mathrm{in}}$ effectively travels unaltered to the reference sphere with radius $f$, where the local wave-vector is tilted towards the geometrical focus under an angle $\theta$ and the CP is preserved for each wave-vector. (\textbf{b}) The beam reflected from the substrate is collected by the same objective and analyzed polarization-resolved in the back focal plane (BFP). The transverse electric and transverse magnetic (TE/TM) polarized fields $E^{\mathrm{TE}}$ and $E^{\mathrm{TM}}$ are aligned with the azimuthal and radial cylindrical unit vectors, respectively. For Brewster's angle $\theta_\mathrm{B}$, the TM component $E^{\mathrm{TM}}$ vanishes. (\textbf{c})  Snapshot in time of CP electric field components in the reflection BFP around the intersection point of y-axis (origin centered in BFP) and Brewster ring. In contrast to the azimuthal TE field, the radial TM component exhibits a phase jump of $\pi$ at $\theta_\mathrm{B}$, due to the properties of the Fresnel reflection coefficient $r^{\mathrm{TM}}$. (\textbf{d}) Projection of (\textbf{c}) onto the y-axis. The projected TE polarized field components on opposite sides of the intersection point possess opposite phase and are inherently $\frac{\pi}{2}$ out of phase with respect to TM in CP light. Consequently, a vortex of topological charge $\ell=\pm1$ emerges, with the sign of the phase charge depending on the handedness of the incoming polarization state. (\textbf{e}) Time evolution of the polarization distribution in the BFP at Brewster's angle for the reflected vortex beam referred to throughout this letter. Note that for reflection the incident CP field on a ring is multiplied by the corresponding Fresnel reflection coefficient, thus there is no TM polarized field $E^{\mathrm{TM}}$ at the ring corresponding to $\theta_\mathrm{B}$.}
		\label{fig:1}
	\end{figure*}   

	\section{\label{sec:theory}Theory} 
	
	Consider a Gaussian beam with its waist $w_0$ coinciding with the BFP of an aplanatic objective with focal length $f$ and $\mathrm{NA} = n\sin(\theta_{max})$, where $n$ is the refractive index of the focusing side, and $\theta_{max}$ denotes the maximum aperture angle. In this case, the incident field is given by
	\begin{equation}
	\vec{E}_\mathrm{in} = \underbrace{\mathrm{E_0} \exp\left( -\frac{f^2 \sin^2{\theta}}{w_0^2} \right)}_{\mathrm{E_{in}}} \vec{e}_\mathrm{in},\quad \theta \leq  \theta_{\mathrm{{max}}} \text{,}
	\label{eq:gauss_in_sph}
	\end{equation}
	with amplitude $\mathrm{E_0}$ and complex polarization vector $\vec{e}_\mathrm{in}$. The spatial extent of the input beam may be characterized by the filling factor $f_0=f\sin\theta_{\mathrm{max}}/w_0$, i.e. the ratio of objective aperture radius to beam waist.\\
	\noindent The lens establishes a link between real-space distribution $\vec{E}_\mathrm{in}(x,y)$ of the paraxial input beam in the BFP and the angular distribution of the focal field $\tilde{\vec{E}}(k_x,k_y)$. The transformation it performs may be illustrated by a reference sphere of radius $f$ around the geometrical focus, to which the beam effectively travels undisturbed. A ray impinging on this reference sphere at a distance $\rho=f \sin \left( \theta\right)$ from the optical axis is refracted such that it propagates towards the geometrical focus under the divergence angle $\theta$, corresponding to a coordinate transformation of the form $(x,y) \rightarrow (-f\frac{k_x}{k},-f\frac{k_y}{k})$~ \cite{novotny2012principles}. The process is schematically illustrated for the case of incident CP in \figref{fig:1}\textbf{a}, whereby upon focusing the polarization is preserved for each wave-vector. Since in our case we consider the BFP in reflection, the lens also performs the back-transformation on the reflected field (cf. \figref{fig:1}\textbf{b}). For deriving the latter, it is convenient to employ the TE/TM polarization basis, with the electric field vector being orthogonal ($\vec{E}^\mathrm{TE}$) or parallel ($\vec{E}^\mathrm{TM}$) to the plane of incidence, respectively. In our cylindrically symmetric focusing system, TE/TM are aligned with the azimuthal and radial unit vectors $\vec{e}_{\phi}$/$\vec{e}_{\rho}$. Consequently, in this basis, the reflected field $\vec{E}_\mathrm{r}$ differs from the incident one only by the factor of the Fresnel reflection coefficients $r^\mathrm{TE/TM}$ \cite{novotny2012principles}: 
	\begin{equation}
	\vec{E}_\mathrm{r} = \underbrace{ \left[r^\mathrm{TM} \mathrm{E_{in}} (\vec{e}_\mathrm{in} \cdot \vec{e}_{\rho})\right]\vec{e}_{\rho}}_{\vec{E}^\mathrm{TM}} + \underbrace{ \left[ r^\mathrm{TE} \mathrm{E_{in}} (\vec{e}_\mathrm{in} \cdot \vec{e}_{\phi})\right]\vec{e}_{\phi}}_{\vec{E}^\mathrm{TE}} \text{.}
	\label{eq:eref}
	\end{equation}
	In the case of a dielectric interface, the reflected TM polarized field $\vec{E}^\mathrm{TM}$ vanishes at Brewster's angle $\theta_B = \arctan\left(n_2/n_1 \right)$, having peculiar consequences for incident CP light ($\vec{e}_\mathrm{in}=\vec{e}_\pm$), as discussed below.
	
	The polarization unit vectors $\vec{e}_{\pm}$ for CP may be written as a phase-delayed superposition of Cartesian unit vectors with $\vec{e}_{\pm} \propto (\vec{e}_x \pm \imath \vec{e}_y)$, where upper and lower sign correspond to left- and right-hand CP (LCP/RCP), respectively. When switching from the circular ($\vec{e}_{\pm}$) to the cylindrical ($\vec{e}_{\rho/\phi}$) basis, the transformation of unit vectors results in  $\vec{e}_\pm \cdot \vec{e}_{\phi} \propto \exp \left( \pm \imath \phi \right)$ for the azimuthal TE polarized field component, i.e. a helical phase profile emerges \cite{bliokh2011spin}. In consequence, the reflected TE polarized field $\vec{E}^\mathrm{TE}$ is an azimuthally polarized vectorial vortex beam with a phase singularity of topological charge $\ell = \pm 1$ on the optical axis. As mentioned, at $\theta_\mathrm{B}$ the TM polarized field $\vec{E}^\mathrm{TM}$ vanishes ($r^\mathrm{TM}=0$) and therefore the purely TE polarized vectorial vortex beam with phase singularity is naturally separated at Brewster's angle. \figref{fig:1}\textbf{e} schematically depicts the time evolution of the polarization distribution in the BFP for this azimuthal vortex beam, clearly showing the presence of singularities and elucidating the origin of a central phase vortex.
	
	Moreover, analysis typically employed in polarization microscopy consists of a projection of the reflected beam onto the Cartesian polarization basis. Remarkably, this results in two phase vortices of charge $\ell=\pm1$, emerging at the respective intersection points of Cartesian axis and Brewster ring. The effect originates from the polarizing properties of the Fresnel reflection coefficients and may be understood intuitively by investigating a snapshot in time of the polarization distribution around one of these points (cf. \figref{fig:1}\textbf{c}). The coefficient $r^\mathrm{TM}$ exhibits a zero crossing at Brewster's angle, which translates to a phase difference of $\pi$ for the radial TM polarized field below and above $\theta_\mathrm{B}$. At the same time, the phase of the azimuthal TE component remains unaltered by $r^\mathrm{TE}$. Projecting the polarization components on the corresponding Cartesian polarization axes results in the field distribution schematically depicted in \figref{fig:1}\textbf{d}. The linear projections of the azimuthal components on opposite sides of the intersection point are $\pi$ out of phase. Together with the inherent $\frac{\pi}{2}$ phase delay of TE with respect to TM in CP light, a helical phase front forms around the intersection point, i.e. an optical vortex of charge $\ell= \pm 1$. As a consequence, phase singularities naturally emerge in reflection in any polarization microscopy setup employing high NA.  
	
	\begin{figure}[!t]
		\centering 
		\includegraphics[width=\columnwidth]{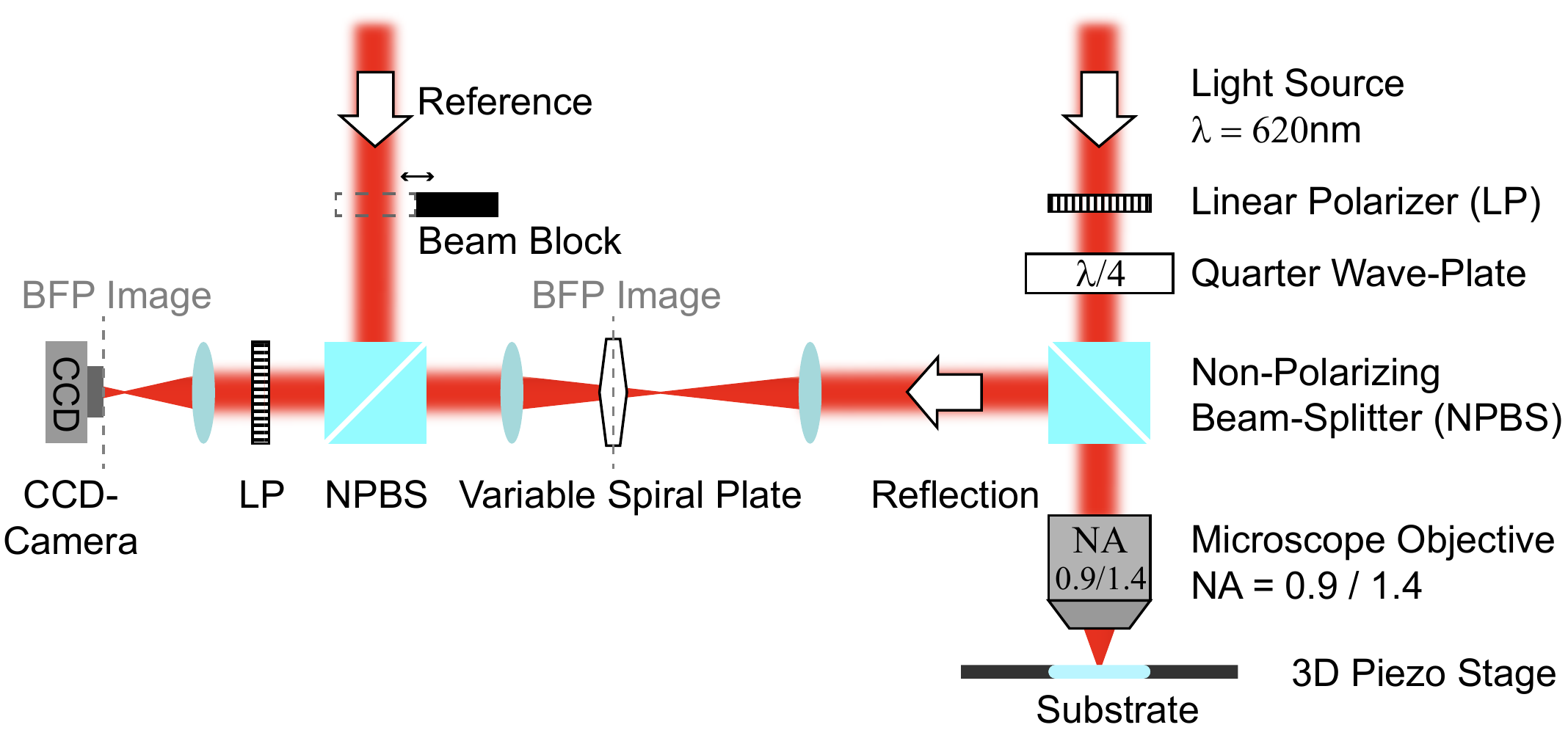} 
		\caption{Simplified sketch of the experimental setup utilized for measuring the BFP in reflection. The BFP is imaged on a variable spiral plate in order to convert azimuthal and radial TE/TM field components to the X/Y laboratory axes. The final linear polarizer before the CCD-camera allows for selecting the intensity distribution of the desired polarization state to be recorded. In addition, the phase profile may be reconstructed via interference with a reference beam. A variation of the setup without variable spiral plate is used to directly project the polarization distribution in the BFP on Cartesian X/Y axes.}
		\label{fig:2}
	\end{figure}   

	\begin{figure}[!t]
	\centering 
	\includegraphics[width=\columnwidth]{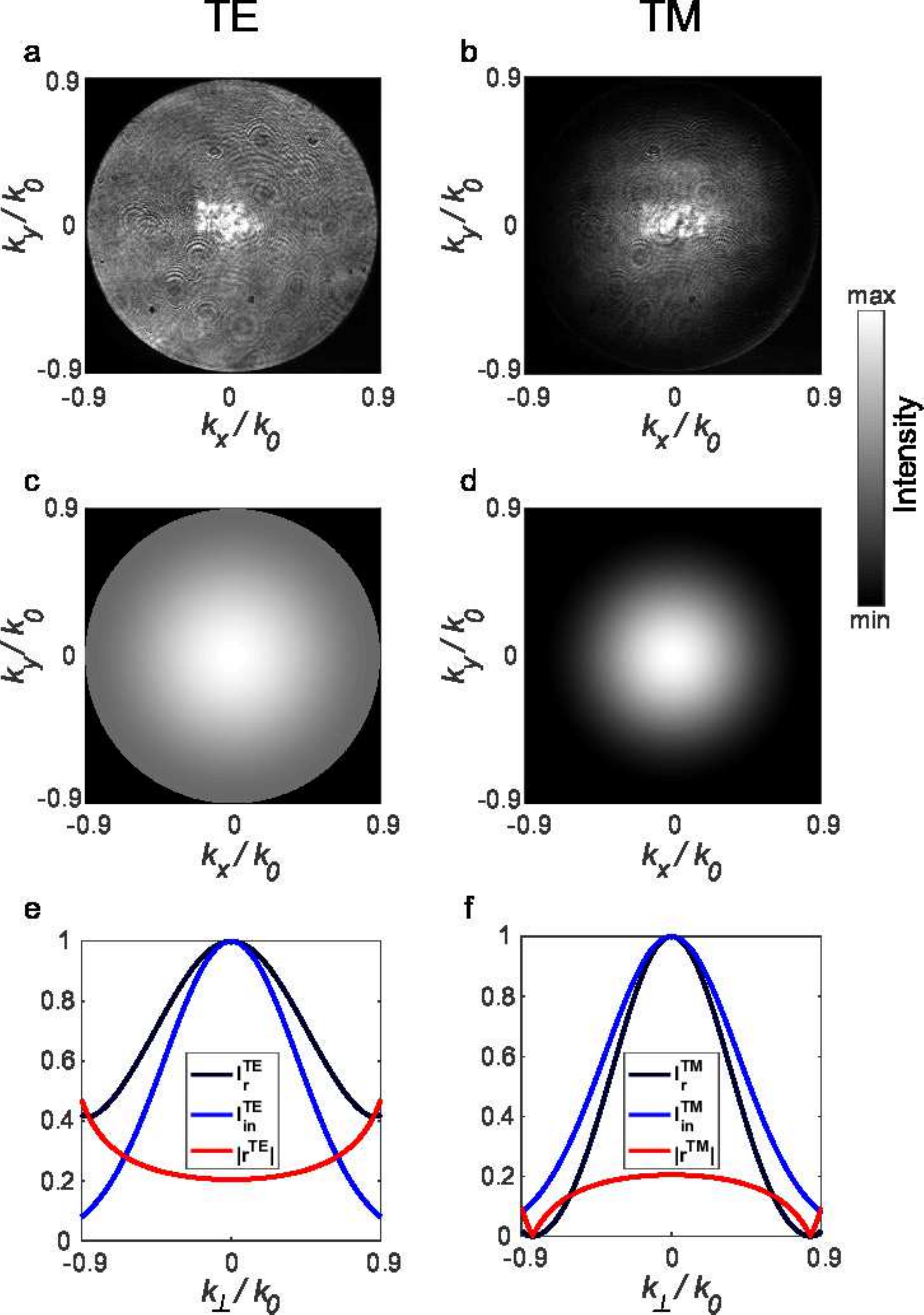} 
	\caption{(\textbf{a}),(\textbf{b}) Experimental and (\textbf{c}),(\textbf{d}) theoretical intensity distributions in the BFP of the $\mathrm{NA} = 0.9$ microscope objective, for reflection of a focused CP Gaussian beam from a BK7 glass substrate. In contrast to the rather homogeneous intensity distribution for TE polarized light (\textbf{a}),(\textbf{c}), the projection on TM (\textbf{b}),(\textbf{d}) shows a null intensity ring with radius corresponding to Brewster's angle $\theta_\mathrm{B}$. (\textbf{e}),(\textbf{f}) Cross-sectional view of the TE/TM projections for the incident and reflected beam's intensity distributions $I_\mathrm{in}$ and $I_\mathrm{r}$ (normalized to their respective maximum), alongside with the absolute value of the corresponding Fresnel reflection coefficient $|r^\mathrm{TE/TM}|$.}
	\label{fig:3}
	\end{figure} 
	
	\section{Setup} 
	
	To demonstrate the emergence of a vectorial vortex beam with phase singularity in reflection at Brewster's angle, we experimentally measure the polarization state and the wavefront of the reflected light in the cylindrical polarization basis. The setup is schematically depicted in \figref{fig:2}. 
	
	We focus a CP Gaussian beam with a wavelength of $\lambda=620\,$nm onto a BK7 glass substrate. The beam is focused tightly by a dry microscope objective of $\mathrm{NA} = 0.9$ ($f_0\approx 0.89$) or an index-matched oil immersion microscope objective of $\mathrm{NA} = 1.4$ ($f_0\approx 0.86$). To decompose the reflected beam into its TE/TM components, we first image the objective's BFP onto a liquid-crystal-based variable spiral plate \cite{Slussarenko:11}, which allows for the generation of radial or azimuthal polarization patterns from incident linear polarization states and vice versa \cite{Cardano:12}. Consequently, the variable spiral plate enables us to convert the radial and azimuthal TE/TM field components resulting from reflection at the planar substrate to the linear X/Y laboratory axes, with a subsequent projection utilizing a rotatable linear polarizer. It is crucial to note here that, following the geometric considerations presented in Sec.~\ref{sec:theory}, at $\theta_\mathrm{B}$ only a purely azimuthally (TE) polarized vectorial vortex beam is reflected, which bears a phase singularity of topological charge $\ell=\pm1$. The emerging vortex beam may be separated e.g. by a suitable ring aperture in the BFP.\\
	For measuring the direct projection of the reflected field on the Cartesian X/Y polarization axes we simply remove the variable spiral plate. Furthermore, interferometric measurements are performed by superposition with a reference beam possessing a planar wave front. The phase profile is successively retrieved from the recorded fringe patterns, following the procedure described by Takeda et al. \cite{Takeda:82}.
	
		\begin{figure}[!t]
		\centering 
		\includegraphics[width=\columnwidth]{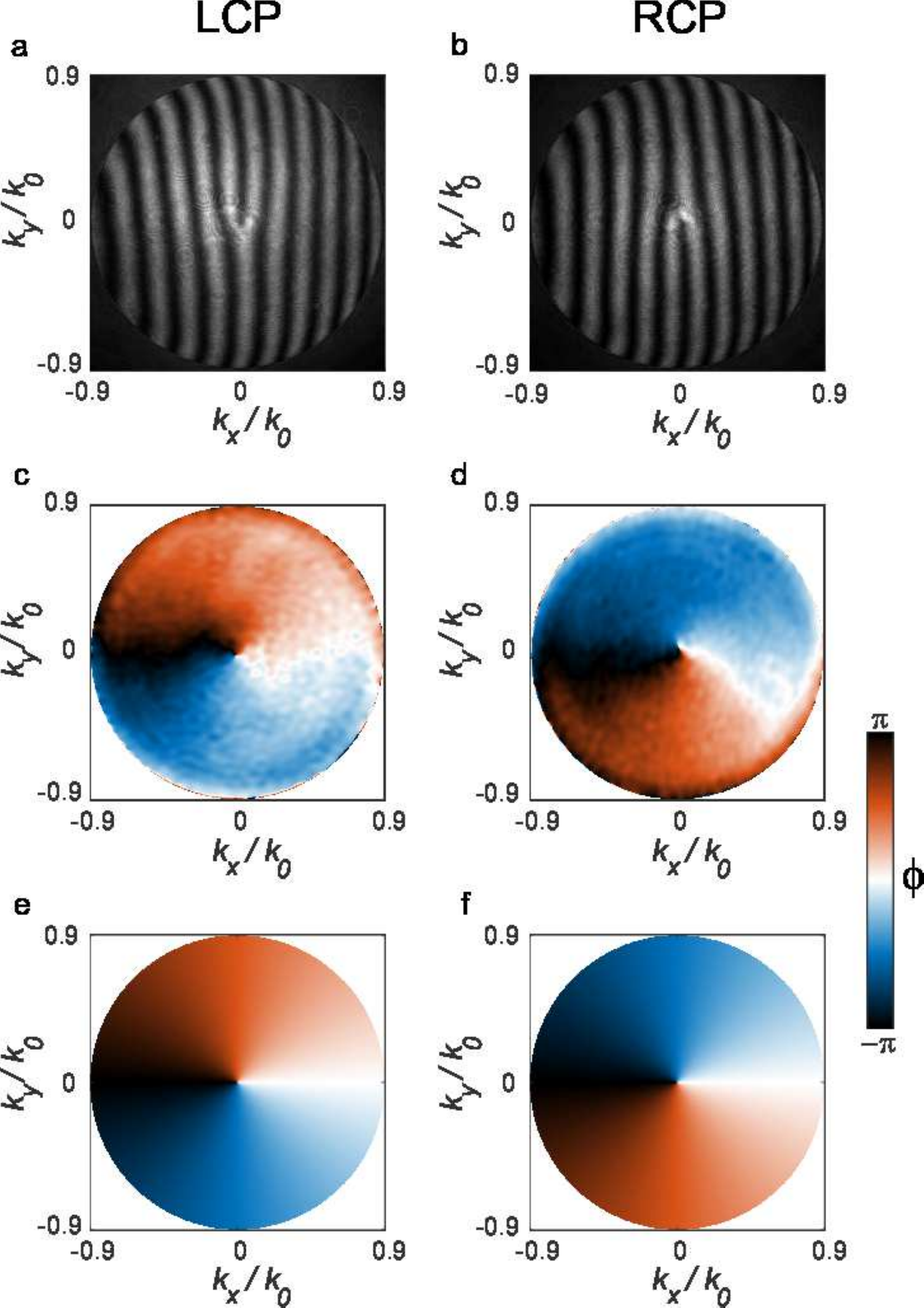} 
		\caption{(\textbf{a}),(\textbf{b}) Interference patterns of TE polarized reflected light with a planar-wavefront reference beam for incident left- and right-hand CP (LCP/RCP) in the BFP of the $\mathrm{NA} = 0.9$ microscope objective, displaying fork holograms of opposite orientations. (\textbf{c}),(\textbf{d}) Corresponding experimental and (\textbf{e}),(\textbf{f}) theoretical phase fronts, verifying the presence of a central singularity surrounded by a helical phase distribution.}
		\label{fig:4}
	\end{figure} 
	
	\begin{figure}[!t]
		\centering 
		\includegraphics[width=\columnwidth]{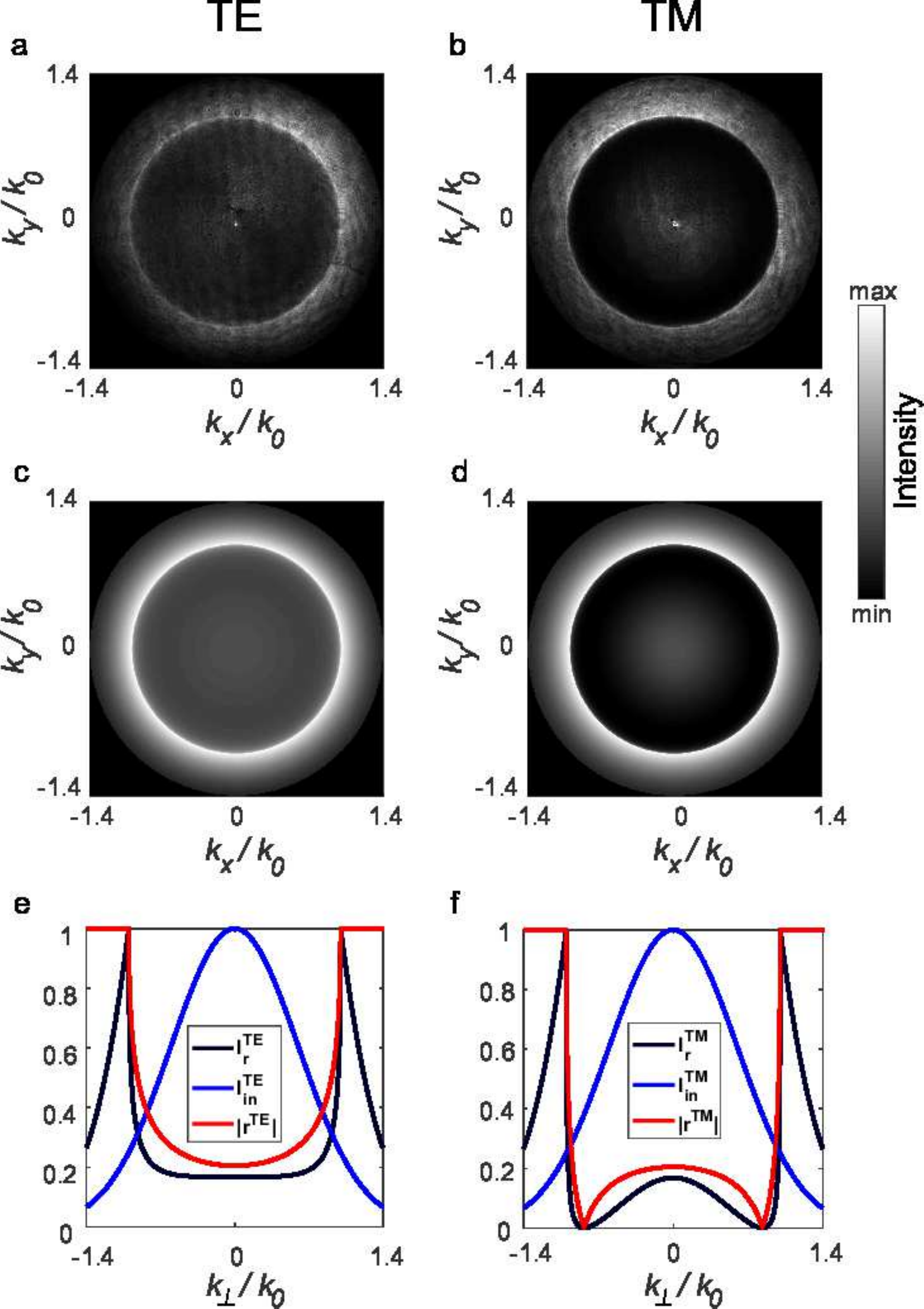} 
		\caption{(\textbf{a}),(\textbf{b}) Experimental and (\textbf{c}),(\textbf{d}) theoretical intensity distributions in the BFP of the $\mathrm{NA} = 1.4$ oil immersion microscope objective, for reflection of a focused CP Gaussian beam from the glass-air interface at the bottom of a BK7 substrate. As for the case with the dry $\mathrm{NA} = 0.9$ objective in \figref{fig:3}, a null intensity ring at Brewster's angle $\theta_\mathrm{B}$ shows up for TM polarized light (\textbf{b}),(\textbf{d}), contrary to the projection on TE polarization (\textbf{a}),(\textbf{c}). Furthermore, the sharp transition at the critical angle for total internal reflection $\theta_c$ and the high intensity above it is prominent in all images. (\textbf{e}),(\textbf{f}) Cross-sectional view of the TE/TM projections for the incident and reflected beam's intensity distributions $I_\mathrm{in}$ and $I_\mathrm{r}$ (normalized to their respective maximum), alongside with the absolute value of the corresponding Fresnel reflection coefficient $|r^\mathrm{TE/TM}|$.}
		\label{fig:5}
	\end{figure} 

	\section{Results and Discussion} 
	
	The theoretically calculated and experimentally recorded intensity distributions of TE/TM polarized light in the reflection BFP of the $\mathrm{NA} = 0.9$ microscope objective are depicted in \figref{fig:3}\textbf{a}-\textbf{d}. In contrast to the TE polarized intensity, TM polarized light exhibits a prominent dark ring towards the outer edges of the BFP \cite{lopez2017refractive}. The different patterns originate from the distinct angular dependence of the Fresnel reflection coefficients $r^{\mathrm{TE/TM}}$, elucidated in the cross-sectional view of the incident and reflected beam's intensity alongside with the evolution of $|r^{\mathrm{TE/TM}}|$ in \figref{fig:3}\textbf{e},\textbf{f}. Since $r^{\mathrm{TM}}$ exhibits a zero crossing at Brewster's angle $\theta_\mathrm{B}$, a null intensity ring with corresponding radius appears in the BFP. 
	
	 When now interfering the TE polarized field with a planar phase front reference beam, a 'fork' hologram with opposite orientation for incident LCP/RCP appears (cf. \figref{fig:4}\textbf{a},\textbf{b}), confirming the presence of a phase vortex with topological charge $\ell=\pm1$. The corresponding reconstructed experimental and theoretical phases are shown in \figref{fig:5}\textbf{c},\textbf{d} and \textbf{e},\textbf{f}, respectively. The phase images validate the natural emergence of a vectorial vortex beam with central phase singularity in the surprisingly common configuration of focused CP light, reflected from a dielectric substrate at Brewster's angle $\theta_\mathrm{B}$.
	
	\begin{figure}[!t]
		\centering 
		\includegraphics[width=\columnwidth]{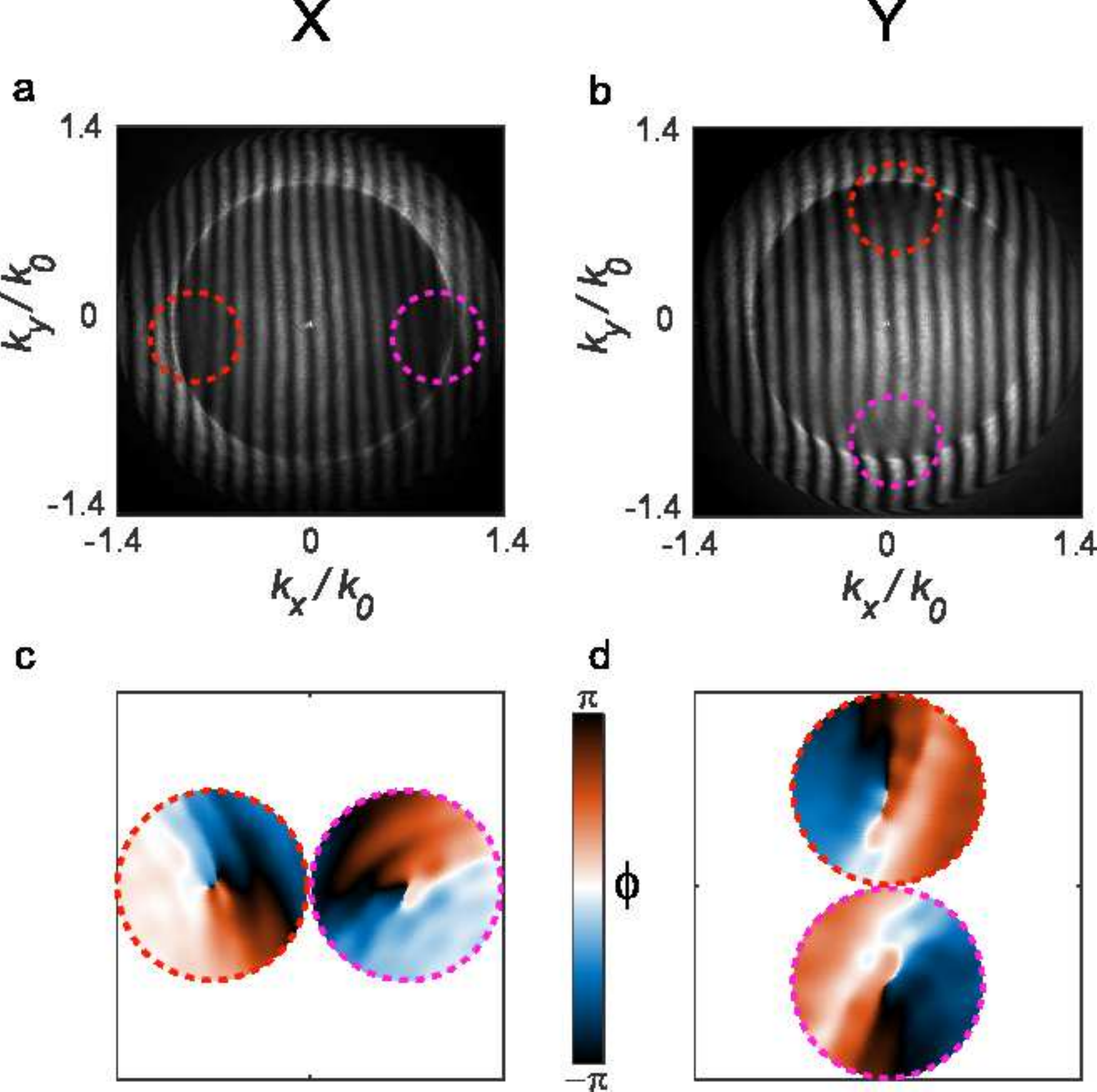} 
		\caption{(\textbf{a}),(\textbf{b}) Interference patterns of X/Y polarized projections of the reflected light with a planar-wavefront reference beam for incident RCP in the BFP of the $\mathrm{NA} = 1.4$ microscope objective. Two horizontally (X) or vertically (Y) aligned forks emerge at the intersection points of Brewster ring and respective Cartesian axis. (\textbf{c}),(\textbf{d}) Phase reconstruction around these points affirms the presence of vortices with topological charge $\ell=\pm1$.}
		\label{fig:6}
	\end{figure} 

	A frequently applied scheme in polarization microscopy utilizes index-matched immersion oil in the focusing path for investigation of specimens on a substrate in a homogeneous environment. Since Brewster's angle also appears for the transition to an optically denser medium, e.g. glass to air, similar effects are observed in this measurement scheme. We present corresponding results  in \figref{fig:5}, using an index-matched oil immersion type objective of $\mathrm{NA} = 1.4$. We benefit from this scheme not only by a broader range of incidence angles, but also by total internal reflection above the critical angle $\theta_c$, which due to its vicinity to $\theta_\mathrm{B}$ increases the visibility of the dark ring at Brewster's angle for TM polarization. The theoretical and experimental intensity distributions for TE/TM nicely reproduce the expected features. Apart from the dark Brewster ring in the TM projection of the BFP, the sharp transition at the critical angle as well as the high intensity above it is evident for both polarizations.	
	
	In the next step, we show that the Brewster effect leads to the emergence of phase vortices even in the ubiquitous case of a polarization projection onto Cartesian axes. Therefore, we remove the variable spiral plate (cf. \figref{fig:2}), resulting in a projection of the reflected beam onto Cartesian X/Y instead of cylindrical TE/TM coordinates with the final linear polarizer. As discussed in Sec.~\ref{sec:theory}, for a projection of the BFP polarization distribution onto X/Y, we expect two phase vortices forming at the intersection points of the Brewster ring and the respective Cartesian axis. Indeed, the interference patterns for incident RCP in \figref{fig:6}\textbf{a},\textbf{b} clearly show the emergence of two horizontally or vertically aligned forks at $\theta_\mathrm{B}$, depending on the Cartesian axis chosen for projection. Likewise, phase reconstruction around both points corroborates the expectation of vortices of topological charge $\ell=\pm1$, as shown in the helical phase profiles in \figref{fig:6}\textbf{c},\textbf{d}. As a result, the presence of these parasitic phase singularities must be considered for any linear polarization projection in reflection from focusing geometries covering Brewster's angle.	
	
	\section{Conclusion} 
	
	In conclusion, we experimentally demonstrated the inevitable emergence of phase singularities and generation of vectorial vortices upon reflection of focused CP light from a dielectric substrate under Brewster's angle. Specifically, we utilized a variable spiral plate to separate the TE and TM polarized field components of the reflected CP Gaussian input beam and directly demonstrated the emergence of an azimuthally polarized vectorial vortex beam with a phase singularity, appearing at Brewster's angle. Additionally, we performed interferometry in reflection to directly reconstruct the phase profile in the BFP. Moreover, we also demonstrated and interpreted the presence of phase singularities for an even simpler measurement scheme, performing a polarization projection of the reflected field distribution in the BFP onto Cartesian axes.\\
	\indent The utilized experimental scheme is so common, especially in the field of polarization microscopy, that our findings have to be considered widely wherever high $\mathrm{NA}$ focusing geometries for phase and polarization sensitive measurements in reflection are employed. Furthermore, our studies provide an experimentally and conceptually straightforward basis for generation of vectorial vortex beams with a phase singularity, which are of great interest in a broad range of applications utilizing exotic states of light. 
	
	\begin{acknowledgments}
		The authors gratefully acknowledge inspiring and fruitful discussions with Pawe\l{} Wo\'z{}niak, Martin Neugebauer and Andrea Aiello.
	\end{acknowledgments}
	
	\bibliographystyle{apsrev4-1}
	\bibliography{brewster_bib}
\end{document}